\documentclass[sigconf, nonacm]{acmart}

\usepackage[ruled,vlined]{algorithm2e}
\usepackage{algpseudocode}
\usepackage{multirow}
\usepackage{stackengine}
\usepackage{makecell}
\usepackage{setspace}
\usepackage{pifont}

\newcommand{\squishlist}{
	\begin{list}{$\bullet$}
		{ \setlength{\itemsep}{1pt}
			\setlength{\parsep}{1pt}
			\setlength{\topsep}{2.5pt}
			\setlength{\partopsep}{0.5pt}
			\setlength{\leftmargin}{1em}
			\setlength{\labelwidth}{1em}
			\setlength{\labelsep}{0.6em}
		}
	}
	\newcommand{\squishend}{
	\end{list}
}

\makeatletter
  \newcommand\figcaption{\def\@captype{figure}\caption}
  \newcommand\tabcaption{\def\@captype{table}\caption}
\makeatother

\newcommand\vldbdoi{XX.XX/XXX.XX}
\newcommand\vldbpages{XXX-XXX}
\newcommand\vldbvolume{17}
\newcommand\vldbissue{12}
\newcommand\vldbyear{2024}
\newcommand\vldbauthors{\authors}
\newcommand\vldbtitle{\shorttitle} 
\newcommand\vldbavailabilityurl{https://github.com/ZJU-DAILY/MQA}
\newcommand\vldbpagestyle{empty} 

\begin{document}
% \title{An Interactive Multi-modal Query Answering System \\ with Retrieval-Augmented Large Language Models}
\title[An Interactive Multi-modal Query Answering System with Retrieval-Augmented Large Language Models]{\texorpdfstring{An Interactive Multi-modal Query Answering System \\with Retrieval-Augmented Large Language Models}{An Interactive Multi-modal Query Answering System with Retrieval-Augmented Large Language Models}}

\author{Mengzhao Wang}
\affiliation{%
  \institution{Zhejiang University}
  % \city{xxx}
  % \state{China}
}
\email{wmzssy@zju.edu.cn}

\author{Haotian Wu}
\affiliation{%
  \institution{Zhejiang University}
  % \city{xxx}
  % \state{China}
}
\email{jxk706060666@gmail.com}

\author{Xiangyu Ke}
\affiliation{%
  \institution{Zhejiang University}
  % \city{xxx}
  % \state{xxx}
}
\email{xiangyu.ke@zju.edu.cn}

\author{Yunjun Gao}
\affiliation{%
  \institution{Zhejiang University}
  % \city{xxx}
  % \state{xxx}
}
\email{gaoyj@zju.edu.cn}

\author{Xiaoliang Xu}
\affiliation{%
  \institution{Hangzhou Dianzi University}
  % \city{xxx}
  % \state{xxx}
}
\email{xxl@hdu.edu.cn}

\author{Lu Chen}
\affiliation{%
  \institution{Zhejiang University}
  % \city{xxx}
  % \state{xxx}
}
\email{luchen@zju.edu.cn}

\begin{abstract}
Retrieval-augmented Large Language Models (LLMs) have reshaped traditional query-answering systems, offering unparalleled user experiences. However, existing retrieval techniques often struggle to handle multi-modal query contexts. In this paper, we present an interactive \underline{\bf M}ulti-modal \underline{\bf Q}uery \underline{\bf A}nswering ({\sf MQA}) system, empowered by our newly developed multi-modal retrieval framework and navigation graph index, integrated with cutting-edge LLMs. It comprises five core components: Data Preprocessing, Vector Representation, Index Construction, Query Execution, and Answer Generation, all orchestrated by a dedicated coordinator to ensure smooth data flow from input to answer generation.
One notable aspect of {\sf MQA} is its utilization of contrastive learning to assess the significance of different modalities, facilitating precise measurement of multi-modal information similarity.
Furthermore, the system achieves efficient retrieval through our advanced navigation graph index, refined using computational pruning techniques.
Another highlight of our system is its pluggable processing framework, allowing seamless integration of embedding models, graph indexes, and LLMs. This flexibility provides users diverse options for gaining insights from their multi-modal knowledge base. A preliminary video introduction of {\sf MQA} is available at \textbf{\textcolor{blue}{\url{https://youtu.be/xvUuo2ZIqWk}}}.

\end{abstract}

\maketitle


\pagestyle{\vldbpagestyle}
\begingroup\small\noindent\raggedright\textbf{PVLDB Reference Format:}\\
\vldbauthors. \vldbtitle. PVLDB, \vldbvolume(\vldbissue): \vldbpages, \vldbyear.\\
\href{https://doi.org/\vldbdoi}{doi:\vldbdoi}
\endgroup

\begingroup
\renewcommand\thefootnote{}\footnote{\noindent
This work is licensed under the Creative Commons BY-NC-ND 4.0 International License. Visit \url{https://creativecommons.org/licenses/by-nc-nd/4.0/} to view a copy of this license. For any use beyond those covered by this license, obtain permission by emailing \href{mailto:info@vldb.org}{info@vldb.org}. Copyright is held by the owner/author(s). Publication rights licensed to the VLDB Endowment. \\
\raggedright Proceedings of the VLDB Endowment, Vol. \vldbvolume, No. \vldbissue\ %
ISSN 2150-8097. \\
\href{https://doi.org/\vldbdoi}{doi:\vldbdoi} \\
}\addtocounter{footnote}{-1}\endgroup

\ifdefempty{\vldbavailabilityurl}{}{
% \vspace{.3cm}
\begingroup\small\noindent\raggedright\textbf{PVLDB Artifact Availability:}\\
The source code, data, and/or other artifacts have been made available at \url{https://github.com/ZJU-DAILY/MQA}.
\endgroup
}

\section{Introduction}
\label{sec: intro}
Query Answering (QA) systems are pivotal in extracting insights from vast knowledge bases, offering intuitive and real-time information retrieval functionality. Traditional QA systems rely on simplistic keyword matching, which lacks semantic comprehension. Recent advancements in Large Language Models (LLMs), such as ChatGPT, have equipped QA systems with sophisticated context-understanding capabilities \cite{FCS_task1}. Major tech companies, like Microsoft and Baidu, have launched their own QA applications based on LLMs, such as New Bing and ERNIE Bot, respectively.
Despite their success, challenges such as hallucinations and outdated knowledge persist in LLM-based QA systems, impacting their performance \cite{huang2023survey}. The introduction of retrieval-augmented LLMs offers a promising solution by incorporating vector search techniques \cite{wang2024survey}. It enables QA systems to provide answers using external knowledge sources, thereby promoting factually consistent and reliable responses \cite{jing2024large}.
However, current retrieval methods, which cater mainly to single-modality situations, struggle in multi-modal QA contexts. Given the growing complexity of user interactions, multi-modal QA has become increasingly important. It allows for a comprehensive understanding and response to queries by considering various input forms, including text, images, etc. To address this, we have recently developed a novel multi-modal retrieval framework, \texttt{MUST} \cite{MUST_ICDE24}, specifically designed to facilitate efficient and accurate multi-modal retrieval. In this demonstration proposal, we focus on the multi-modal QA task, leveraging the capabilities of \texttt{MUST} alongside LLMs.

\begin{figure}
  \centering
  \setlength{\abovecaptionskip}{0.1cm}
  \setlength{\belowcaptionskip}{-0.3cm}
  \includegraphics[width=\linewidth]{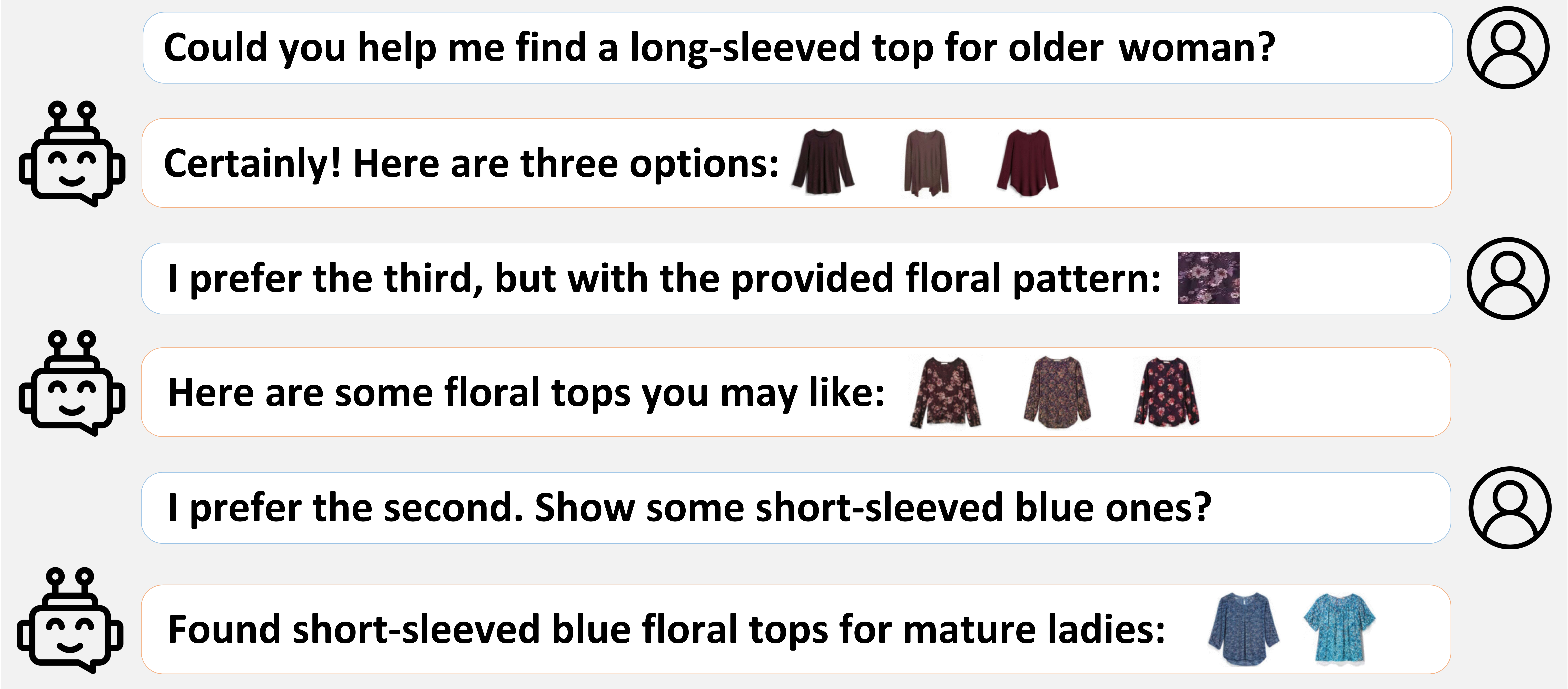}
  \caption{An example of multi-modal QA.}
  \label{fig:problem example}
  \vspace{-0.2cm}
\end{figure}

Figure \ref{fig:problem example} illustrates how users can engage in multi-round dialogues with a QA system, incorporating multi-modal information. For instance, a user may initially submit an inquiry for a ``\textsf{long-sleeved top for older women}'' in text or audio form. Subsequently, based on the images returned in response, the user can choose a preference and suggest alterations, such as adding a ``\textsf{floral pattern}'' to the selected image. This interaction process continues until the user is satisfied with the outcomes.
Multi-modal QA diverges from the conventional text-only QA model by offering more complex and tailored results, significantly enhancing the user experience \cite{DelmasRCL22}.
Despite the demonstrated proficiency of advanced LLMs in understanding multi-modal content (such as GPT-4 \cite{gpt4}), the retrieval techniques currently employed by LLMs are inadequate for capturing multi-modal contexts. This is evident in two prevalent multi-modal retrieval frameworks, Multi-streamed Retrieval (\texttt{MR}) \cite{Milvus_sigmod2021} and Joint Embedding (\texttt{JE}) \cite{DelmasRCL22}.
These frameworks either combine the results of individual vector searches for each modality or carry out a single-channel vector search by jointly encoding all modalities.
However, as empirically confirmed in our \texttt{MUST} paper \cite{MUST_ICDE24}, both baselines exhibit limitations in efficiency and accuracy due to their inability to consider the varying importance of fusing information across modalities and the absence of a dedicated indexing and search method for multi-modal data.

To address the aforementioned challenge, we develop an interactive \underline{\bf M}ulti-modal \underline{\bf Q}uery \underline{\bf A}nswering ({\sf MQA}) system based on our meticulously designed multi-modal retrieval framework \cite{MUST_ICDE24} and navigation graph index \cite{Starling}, integrated with cutting-edge LLMs.
This system offers advanced multi-modal data interaction capabilities via an intuitive interface. The key features are outlined below:

\vspace{0.27em}
\noindent\textbf{User-friendliness.}
{\sf MQA} presents an intuitive interface that accommodates multiple modalities, such as text and images, facilitating customizable searching within a multi-modal knowledge base.
A key aspect of the system is its iterative refinement process, empowering users to fine-tune initial results through ongoing dialogue. In image retrieval, {\sf MQA} transforms descriptive text into visuals, establishing a feedback loop where additional details refine the outcomes, guiding the system towards more relevant outputs, even with vague initial inputs. Moreover, our system integrates advanced LLMs, enhancing the intelligence of the interaction process.

\vspace{0.27em}
\noindent\textbf{Accuracy.}
{\sf MQA} improves query accuracy by utilizing an innovative multi-vector representation technique across multi-modal data. It encodes objects and queries using standalone unimodal encoders or a complex multi-modal encoder, achieving comprehensive vectorized representation. 
Additionally, the method is refined by our unique, effective vector weight learning model \cite{MUST_ICDE24}, capturing individual modality importance through contrastive learning for better similarity evaluations. This empowers users to meticulously adjust searches, hence aligning results with their distinct preferences.

\vspace{0.27em}
\noindent\textbf{Flexibility.}
{\sf MQA} demonstrates flexibility, embracing a wide range of encoders, weight configurations, index algorithms, and retrieval frameworks. Firstly, it supports seamless encoder integration, such as LSTM, ResNet, and CLIP \cite{MUST_ICDE24}. Additionally, it allows tailored weight adjustments using intrinsic vector weight learning or user-specific inputs for search refinement. Moreover, it enables index deployment via an intricate navigation graph framework, effortlessly incorporating existing components. Lastly, it accommodates various retrieval frameworks—\texttt{MR}, \texttt{JE}, and \texttt{MUST}—as well as advanced LLMs like GPT-4, fostering a robust ecosystem for multi-modal QA.

\vspace{0.27em}
\noindent\textbf{Scalability.}
To meet efficiency requirements in large-scale data retrieval, {\sf MQA} employs an advanced navigation graph index designed for multi-modal data \cite{MUST_ICDE24}. This approach assigns multiple vectors per object to a unified index, capturing object similarities. The resulting structure, with vertices representing objects and edges reflecting similarity, narrows the search space, ensuring direct retrieval with minimal traversal. Additionally, {\sf MQA} optimizes both the index structure and multi-vector computations to enhance scalability over a vast knowledge base.
\section{System Overview}
\label{sec: overview}
\begin{figure}
  \centering
  \setlength{\abovecaptionskip}{0.1cm}
  \setlength{\belowcaptionskip}{-0.3cm}
  \includegraphics[width=\linewidth]{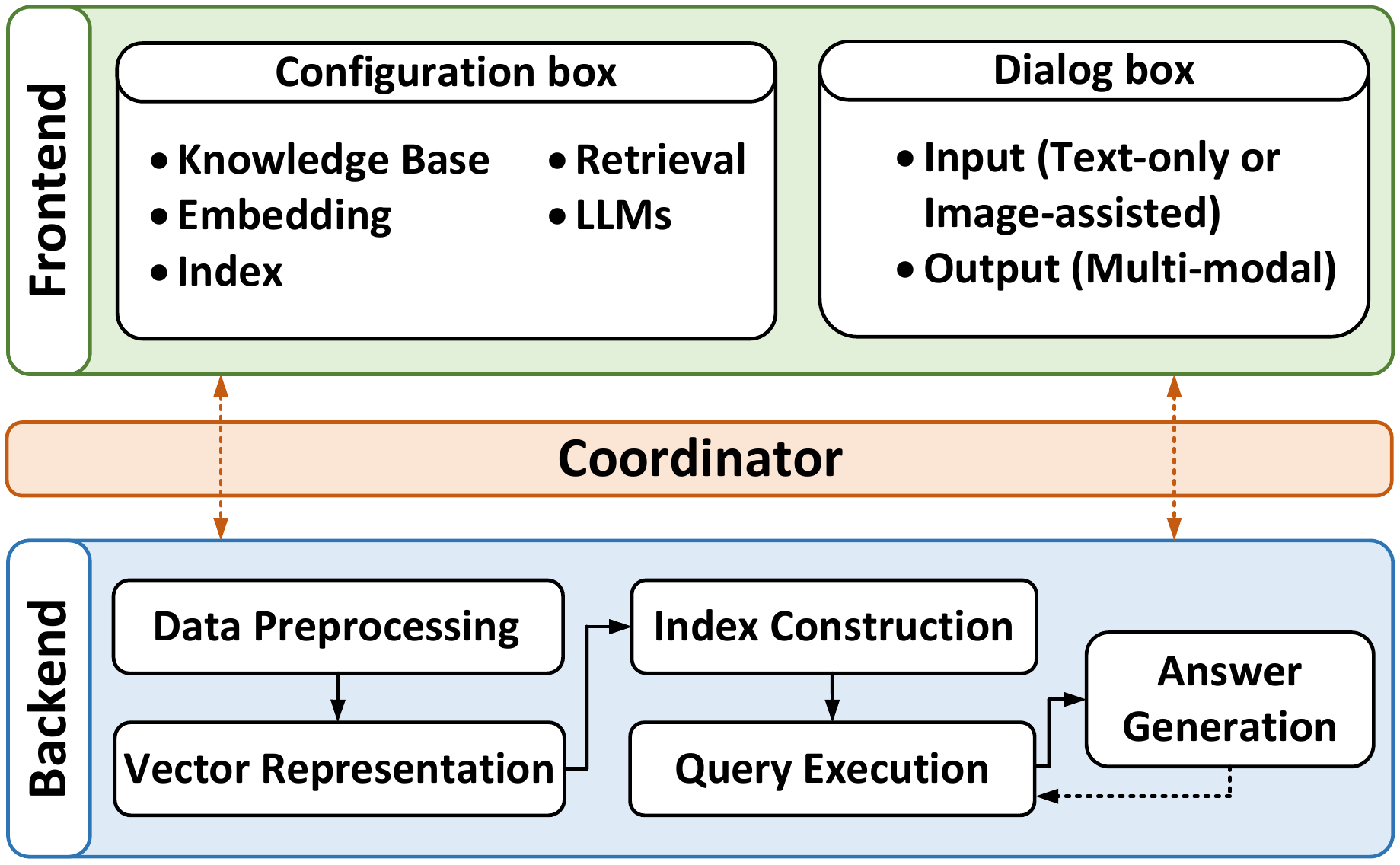}
  \caption{The overall architecture of {\sf MQA}.}
  \label{fig:overview}
  \vspace{-0.1cm}
\end{figure}

Figure \ref{fig:overview} illustrates the system architecture of {\sf MQA}, including both the frontend and backend. The frontend offers a user interface for configuring the knowledge bases, embedding models, index algorithms, retrieval parameters, and LLMs. It also offers a dialogue box for users to engage in multi-modal QA interactions. The backend handles data preprocessing, vector representation, index construction, query execution, and answer generation. A coordinator supervises all components' operations and their data transitions.

\begin{figure*}[!th]
\setlength{\abovecaptionskip}{0cm}
\setstretch{0.9}
\fontsize{8pt}{4mm}\selectfont
\begin{minipage}{0.752\textwidth}
  \setlength{\abovecaptionskip}{0cm}
  \setlength{\belowcaptionskip}{0.1cm}
  \centering
  \footnotesize
  \stackunder[0.8pt]{\includegraphics[scale=0.152]{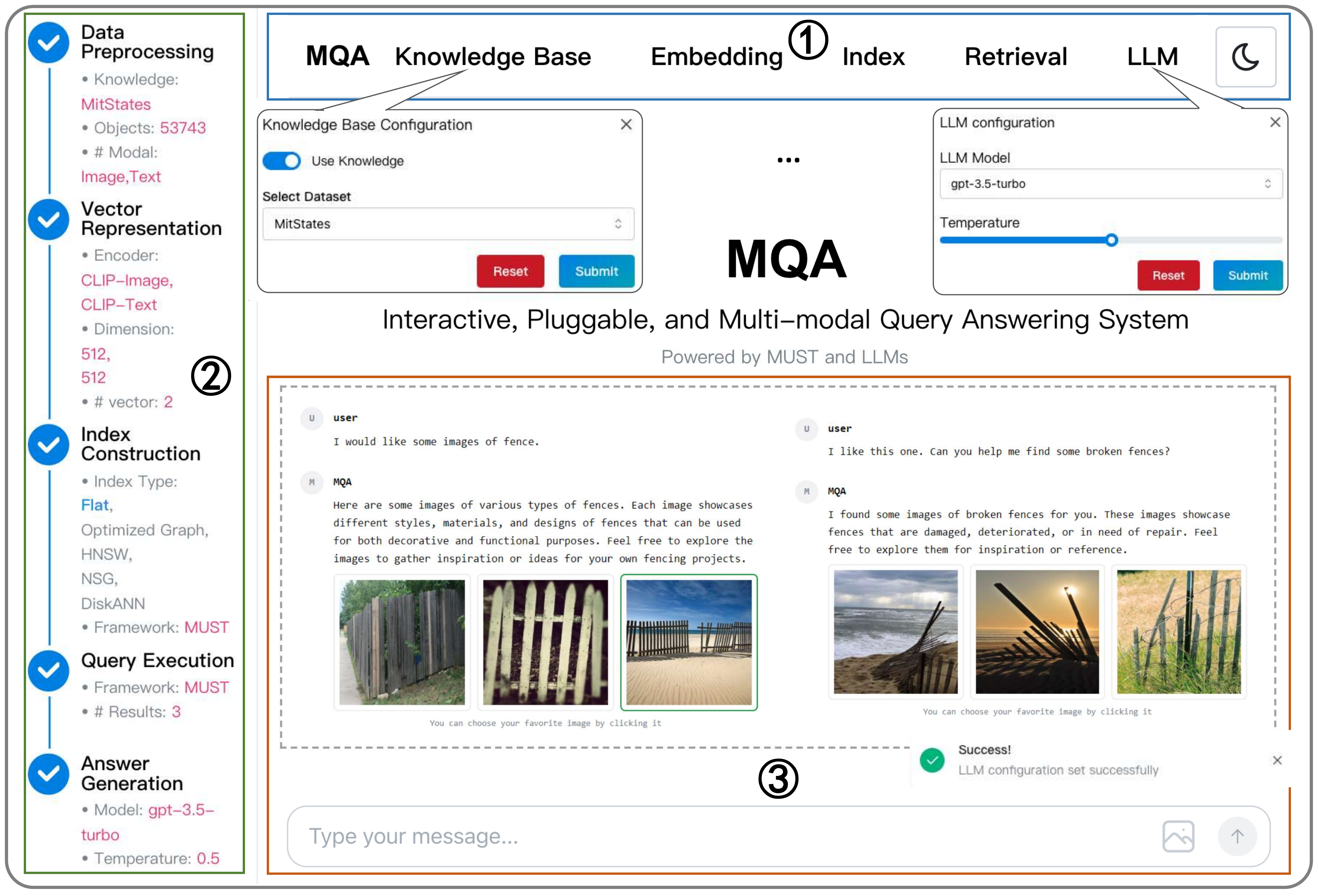}}{}
  \caption{User interface of {\sf MQA}.}
  \label{fig:ui}
\end{minipage}
\begin{minipage}{0.242\textwidth}
  \setlength{\abovecaptionskip}{0.1cm}
  \centering
  \footnotesize
  \stackunder[0.75pt]{\includegraphics[scale=0.051]{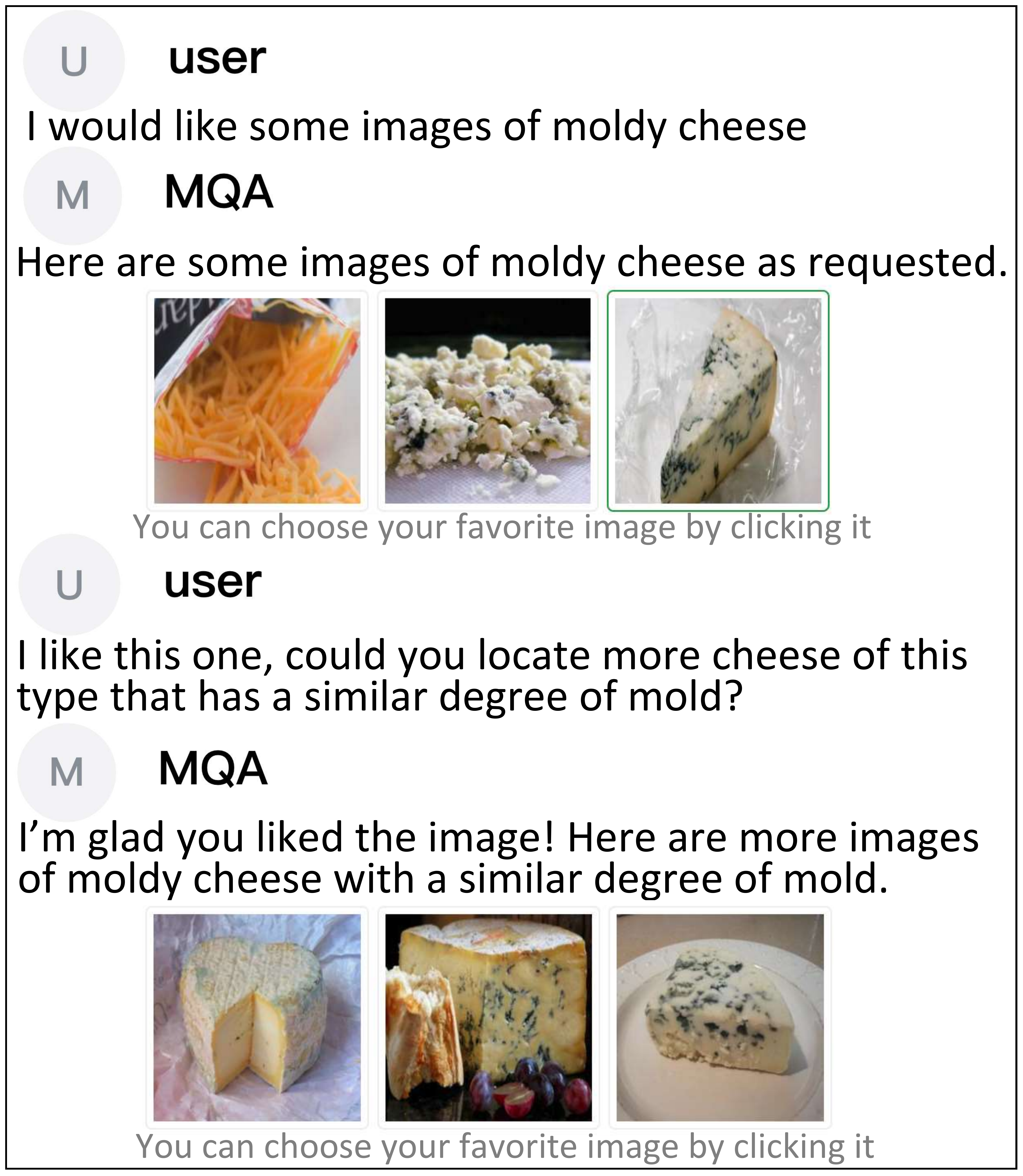}}{(a) Text-only input}
  \\
  \stackunder[0.75pt]{\includegraphics[scale=0.051]{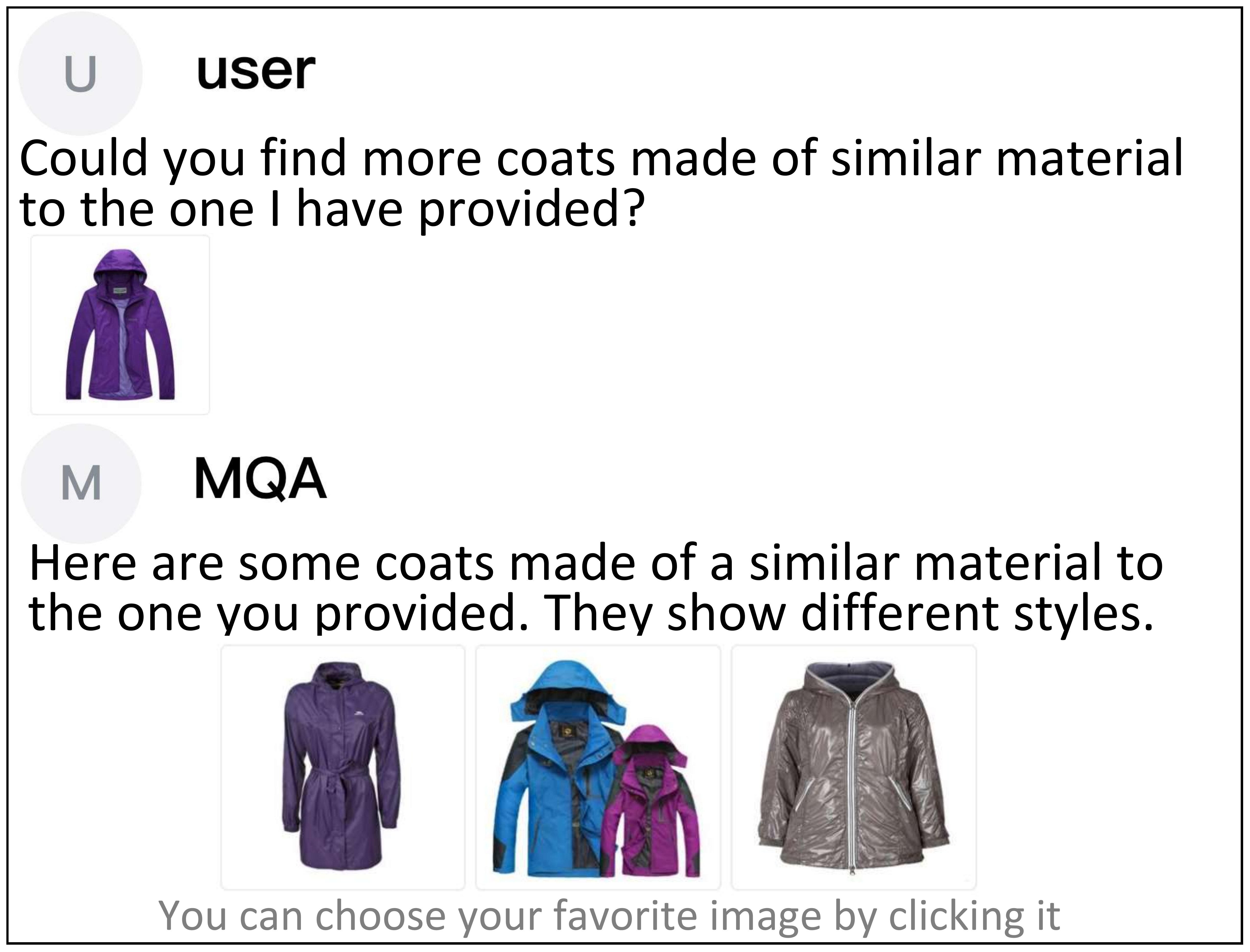}}{(b) Image-assisted input}
  \hspace{-0.2cm}
  \caption{Interaction examples.}
  \label{fig:two example}
\end{minipage}
\vspace{-0.5cm}
\end{figure*}

\vspace{0.2em}
\noindent\textbf{Data Preprocessing.}
This component integrates a multi-modal knowledge base into {\sf MQA}. Data is stored as an object collection with unique IDs for indexing, allowing an expansive data representation. For instance, a movie's film, poster, and synopsis can be stored as a singular object with multiple modalities. Once ingested, data becomes usable within {\sf MQA}, which enables users to leverage the system's embedding and indexing techniques for more effective data storage and retrieval.
Importantly, external knowledge ingestion is optional, and disabling it means {\sf MQA} relies solely on chosen LLMs for responses. This setting is adjustable in the knowledge base section of the configuration box (the frontend of Figure \ref{fig:overview}).

\vspace{0.2em}
\noindent\textbf{Vector Representation.}
This module converts multi-modal objects into vectorized forms, establishing a standardized mathematical expression \cite{MUST_ICDE24}. The frontend embedding configuration includes a universal vector support function, endorsing diverse libraries and models, such as OpenAI CLIP.
{\sf MQA} demonstrates remarkable versatility in representing an object or query using high-dimensional vectors derived from a range of encoders.
Notably, {\sf MQA} introduces a vector weight learning model to discern the importances of different modalities for similarity measurement between objects \cite{MUST_ICDE24}.
The learning process adaptively adjusts weights to reflect different modalities' importances in similarity measurement.
Ultimately, {\sf MQA} outputs the learned weights for vector concatenation, accomplishing a comprehensive representation of multi-modal data.

\vspace{0.2em}
\noindent\textbf{Index Construction.}
The index construction component builds a unified navigation graph index based on objects' multi-vector representation, utilizing modality weights from the vector representation component. The navigation graph corresponds each vertex to an object and forges connections between object pairs through edges denoting object similarity. This index narrows the search space, directing the query to the target object by probing a subset of the collection.
We propose a general pipeline for constructing fine-grained navigation graphs on CGraph\footnote{\url{https://github.com/ChunelFeng/CGraph}}, a cross-platform Directed Acyclic Graph (DAG) framework. The pipeline consists of five flexible parts, allowing any current navigation graph to be decomposed and smoothly integrated into {\sf MQA}.
Furthermore, we incorporate components from several state-of-the-art algorithms in the context of concatenated vectors, resulting in a novel indexing algorithm.
On the configuration box’s index item, users can modify existing navigation graphs (e.g., NSG, HNSW, DiskANN, Starling \cite{Starling}) or initiate custom graphs via the backend API.

\vspace{0.2em}
\noindent\textbf{Query Execution.}
The query execution component navigates efficiently to relevant contexts with user queries using multi-modal vector search methods. Upon receiving a query, {\sf MQA} launches a merging-free search across modalities within the navigation graph. The query is projected into a high-dimensional vector space, where multi-modal objects are located, forming a query point.
{\sf MQA} traverses the graph, starting at a random or fixed vertex, and explores neighboring vertices closer to the query point. This iterative search terminates when no closer vertex is discovered. In this process, distances are calculated via incremental scanning, enhancing efficiency by circumventing unnecessary calculations.
Notably, any previous outcome can be chosen to augment the current user query input (as indicated by the dotted arrow in the backend of Figure \ref{fig:overview}), promoting an intelligent multi-modal search procedure.
Users can modify retrieval settings, like result count, framework, and modality weights at the query point, in the frontend’s configuration box.

\vspace{0.2em}
\noindent\textbf{Answer Generation.}
This component formulates responses from retrieved results and the user query context. Beyond sourcing relevant information from the knowledge base, {\sf MQA} generates natural, conversational replies, enhancing user interactions. The output includes additional details like preference markers. Users have the flexibility to select from various LLMs in the configuration box. When an LLM is accessible, it works in coordination with the query execution module to handle queries and responses. The user’s query is simultaneously dispatched to both the query execution module and the LLM as a prompt. The search results from the query execution module are then redirected to the LLM. The final user response is a summary from the LLM. In the absence of an available LLM, users can still carry out a multi-modal QA procedure through direct engagement with the query execution module.

\vspace{0.2em}
\noindent\textbf{Coordinator.}
The coordinator serves as the system's central nexus, supervising all component operations and facilitating smooth data transition across the system. Both the frontend and backend exclusively interact with the coordinator, which functions as a conduit between them, as demonstrated by two-way arrows in Figure \ref{fig:overview}. This arrangement fosters a streamlined and efficient codebase where API endpoints engage with a single reference point within {\sf MQA}.
\section{Demonstration}
\begin{figure}
  \centering
  \setlength{\abovecaptionskip}{0cm}
  \setlength{\belowcaptionskip}{-0.3cm}
  \includegraphics[width=\linewidth]{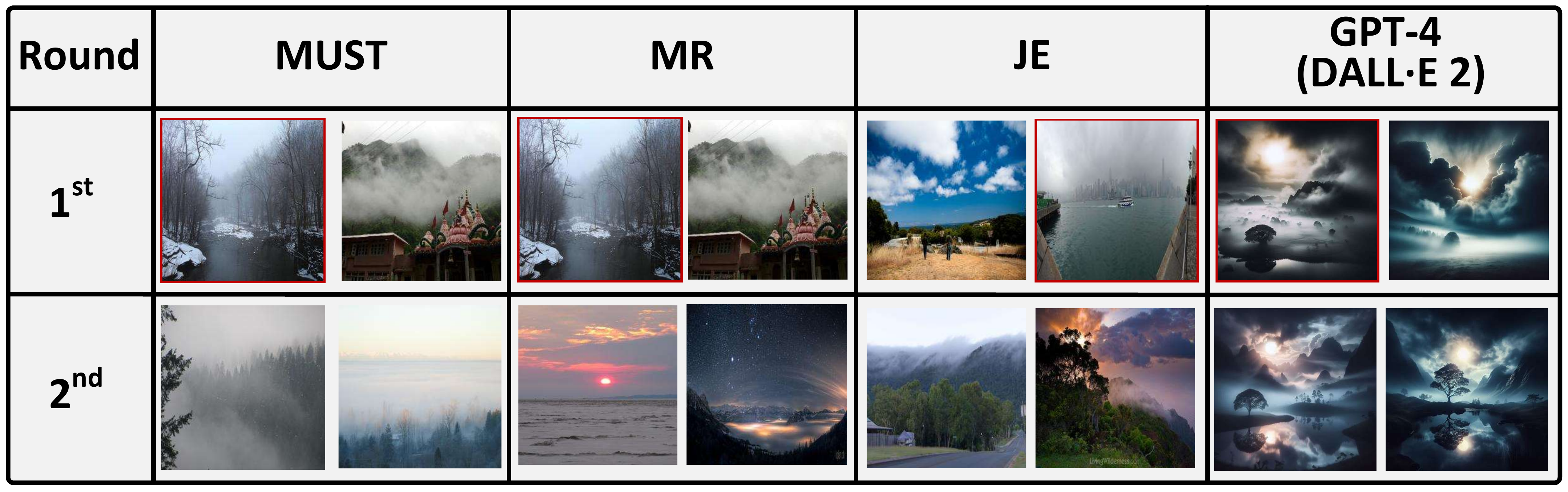}
  \caption{Results from two-round response using different retrieval frameworks, with the user's choice marked in red.}
  \label{fig:comparison}
  % \vspace{-0.1cm}
\end{figure}

{\sf MQA}'s backend is built on the Flask framework, coupled with a user-centric frontend developed with React, Remix, and Mantine that ensures an intuitive user experience filled with interactive features. The demonstration begins with an introductory tour that displays the working panels of {\sf MQA} (refer to Figure \ref{fig:ui}), providing a foundational understanding of user-system interaction. Subsequently, we delve into a hands-on exploration of {\sf MQA}'s features from the user's perspective (see Figure \ref{fig:two example}). To exhibit our techniques' superiority, we provide comparative results with three baseline methods within {\sf MQA} for identical query inputs (see Figure \ref{fig:comparison}).

\vspace{0.2em}
\noindent\textbf{Working Panels.}
As depicted in Figure \ref{fig:ui}, the {\sf MQA} system encompasses three panels for user interaction: \ding{172} configuration, \ding{173} status monitoring, and \ding{174} query-answering (QA) engagement.

% \vspace{0.1em}
\underline{\ding{172} Configuration Panel.}
This interface empowers users to explore the system's features. It facilitates the selection of domain-specific knowledge bases, customizing the search results. Upon a user's selection of a knowledge base, the system initiates data loading, establishing the necessary knowledge base for retrieval. Through embedding options, various encoders can be adjusted for multi-modal data embedding, a critical process for transforming multi-modal data into a vectorized format. The option to activate vector weight learning customizes index construction and retrieval processes with a unique weighting mechanism for the generated vectors. Indexing configurations provide choices on methods and parameters, setting the stage for efficient data retrieval. Additionally, retrieval settings allow users to dictate the retrieval framework and the size of the result set. LLM options present a selection of models and control over output variability via temperature settings. Configuration feedback is relayed through a pop-up box (located in the bottom right corner of Figure \ref{fig:ui}), ensuring users are informed of the system's setup status.

% \vspace{0.1em}
\underline{\ding{173} Status Monitoring Panel.}
This panel displays a comprehensive overview of the system's workflow, from data input to output, providing real-time updates. Milestones such as data preprocessing, vector representation, and index construction are visibly tracked with tick marks and relevant details, encompassing encoder details, modal counts, vector dimensions, index types, retrieval frameworks, and LLM specifics. This feature ensures that users can verify and assess their custom settings at a glance.

% \vspace{0.1em}
\underline{\ding{174} QA Panel.}
Serving as the interface for query submissions, this module accepts both textual input and image uploads from users to form a multi-modal query. It promptly returns relevant multi-modal information, using an optimized retrieval mechanism guided by LLM to ensure context accuracy. Users can fine-tune these results by providing additional input, leading to tailored content. Additionally, this module includes a demonstration example, providing a practical reference to enhance user interaction.

\vspace{0.2em}
\noindent\textbf{Interaction Scenarios.}
{\sf MQA} enhances user engagement through two primary interaction scenarios, illustrated in Figure \ref{fig:two example}, with detailed configurations displayed in the corresponding status monitoring panel (Figure \ref{fig:ui}).
(a) Text-only input: In the absence of a reference image, users can initiate a search using a text description, such as ``\textsf{I would like some images of moldy cheese}''. The system responds by retrieving images that match the description. Users can select a preferred image (by clicking) and refine their request, possibly by adding ``\textsf{I like this one, could you locate more cheese of this type that has a similar degree of mold?}''. {\sf MQA} iteratively refines the search based on user feedback until the user is content with the images retrieved.
(b) Image-assisted input: When a refer- ence image is available, users upload it and describe their specific requirements, for example, ``\textsf{Could you find more coats made of similar material to the one I have provided?}''. The {\sf MQA} system analyzes the visual and textual information, delivering images that align with both the reference image and the textual specifications. Users can further interact by selecting a preferred result for more personalized searches.

\vspace{0.2em}
\noindent\textbf{Comparative Analysis.}
In {\sf MQA}, we compare the outcomes of various retrieval frameworks under identical query conditions. We adjust the index and retrieval techniques by the configuration panel, with Figure \ref{fig:comparison} illustrating the system's two-round response. The user begins with a textual request: ``\textsf{Could you assist me in finding images of foggy clouds?}''. Subsequently, upon specifying a preference ``\textsf{I like this one, could you provide more similar images of foggy clouds?}'', {\sf MQA} returns results based on the selected image and new user feedback. \texttt{MUST} consistently delivers optimal results in both rounds. In contrast, the \texttt{JE} framework underperforms, initially presenting an irrelevant image, followed by two images that do not align with the user's selection. Although \texttt{MR} initially matches \texttt{MUST}'s results for text-only input, it fails to maintain alignment with the multi-modal inputs in the subsequent round. GPT-4 (DALL·E 2), lacking multi-modal retrieval configurations, generates synthetic images that miss a touch of realism.

%we identify the main bottleneck of MVSS

%Computation reuse: optimize the complex combinations using simple combinations

\begin{acks}
 This work was supported in part by the NSFC under Grants No. (62025206, U23A20296, and 62102351), Zhejiang Province's ``Lingyan'' R\&D Project under Grant No. 2024C01259, and Ningbo Yongjiang Talent Introduction Programme (2022A-237-G). Yunjun Gao is the corresponding author of the work. 
\end{acks}

%\clearpage

\bibliographystyle{ACM-Reference-Format}
\bibliography{myref}

\end{document}